\documentclass{aa501}
\usepackage{graphicx}

\begin{document}

\title{Discovery of two infrared supernovae:\\
a new window on the SN
search\thanks{Based on observations obtained at the ESO-NTT in
La Silla, at the ESO-VLT in Paranal and at
 the Italian Telescopio
Nazionale Galileo (TNG) operated on the island of La Palma by the Centro
Galileo Galilei of the CNAA (Consorzio Nazionale per l'Astronomia e
l'Astrofisica) at the Spanish Observatorio del
Roque de los Muchachos of the Instituto de Astrofisica de Canarias.}}

\subtitle{}

\author{R. Maiolino\inst{1} \and L. Vanzi\inst{2} \and F. Mannucci\inst{3}
  \and G. Cresci\inst{4} \and  F. Ghinassi\inst{5} \and M. Della Valle\inst{1}}
\institute{Osservatorio Astrofisico di Arcetri, Largo E. Fermi 5, Firenze, Italy
\and European Southern Observatory (ESO), Alonso de Cordova 3107,
 Santiago, Chile
\and C.A.I.S.M.I.-C.N.R., Largo E. Fermi 5, Firenze, Italy
\and Dipartimento di Astronomia, Universit\`a di Firenze,
    Largo E. Fermi 5, Firenze, Italy
\and Telescopio Nazionale Galileo, Aptdo de Correos, 565, 38700 Santa
 Cruz de La Palma, Canary Islands, Spain}

\offprints{R. Maiolino}

\date{Received  ; accepted}

\abstract{We report the discovery of two supernovae (SN 1999gw and SN 2001db)
obtained
within the framework of an infrared (2.1$\mu$m) monitoring campaign of
Luminous Infrared Galaxies, aimed at detecting obscured
supernovae.  SN 2001db, extinguished by $\rm A_V \approx 5.5$ mag,
is the first supernova discovered in the infrared which has received
the spectroscopic confirmation. This result highlights the power of
infrared monitoring in detecting obscured SNe and indicates
that optical surveys are probably missing a significant fraction of
SNe, especially in obscured systems such as starburst galaxies.
The preliminary estimate of SN rate in LIRG galaxies is about an
order of magnitude higher than that expected from optical surveys.
 }

\authorrunning{Maiolino et al.}
\titlerunning{Infrared SNe}

\maketitle

\begin{figure*}[t!]
\centering
\includegraphics[width=12truecm,angle=0]{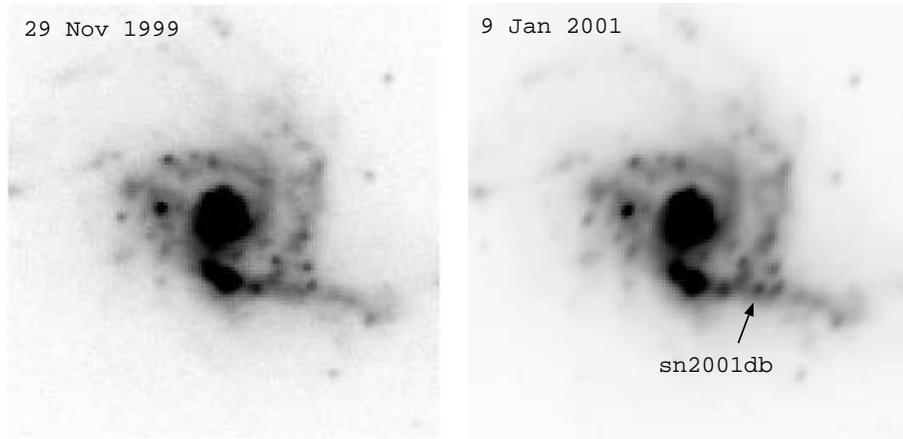}
\caption{Left: Ks-band image of NGC3256 obtained with SOFI in November 1999.
Right: Ks-band image obtained in January 2001 where SN2001db has
been detected.
Both images are $35''\times 35''$ in size.
}
\end{figure*}

\begin{figure*}[htb!]
\centering
\includegraphics[width=8truecm,angle=0]{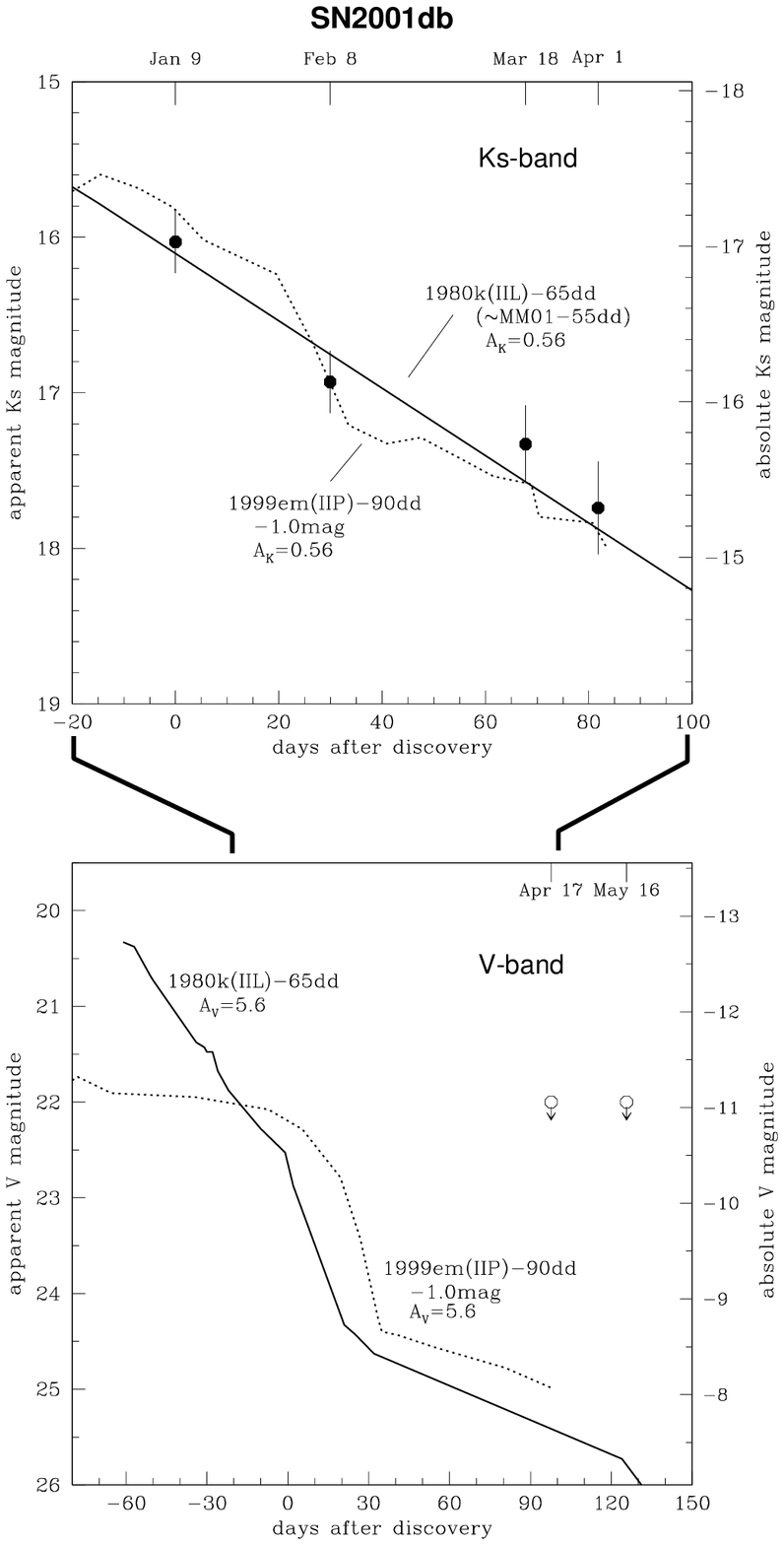}
\caption{{\sl Upper panel:} Ks-band light curve of the SN2001db.
The solid line is the light curve of the SN1980k (type IIL) whose epoch
has been shifted by 65 days (i.e. B band peak at --65 days),
and extinguished by A$_K =0.56$ mag (as
derived in sect.2.2).
This is nearly the same as the average light
curve found by Mattila \& Meikle (\cite{mattila}, MM01),
shifted by 55 days.
The dotted line is
the light curve of SN1999em, whose epoch is shifted by 90 days and whose
magnitude is made brighter by 1 mag (besides the extinction of A$_K =0.55$).
The shift of the template light curve were determined to best fit
the observed light curve.
{\sl Lower panel:}
Upper limits on the V magnitude for SN2001db compared with the
V-band light curves corresponding to those in the upper panel
(extinguished by A$_V =5.6$ mag).}
\end{figure*}

\section{Introduction}

Some recent surveys aimed at detecting SNe (in the optical) have found
that starburst galaxies do not show evidence for an enhanced SN rate,
when the rate is normalized to the B luminosity of the galaxies, i.e.
starburst and quiescent have the same SN rate in terms of
SNu\footnote{$\rm 1SNu = 1SN(100yr)^{-1}(10^{10}L_{\odot}^B)^{-1}$}
(Richmond et al. \cite{richmond}).
This result is
puzzling, since the presence of a burst of star formation should boost
the SN rate much more than the optical luminosity (in addition to the
fact that the optical luminosity is diluted by the quiescent,
evolved stellar population).
One possible explanation for the shortage of SNe in star
forming galaxies
is that most SNe are obscured by the
large amount of dust which generally affects starburst systems.
In the following we will focus on the latter scenario.

A survey for SNe in the infrared, where dust extinction is greatly reduced,
would allow one
 to detect the obscured SNe which are possibly missed in optical
monitoring campaigns. This strategy was first proposed by Van Buren \&
Norman (\cite{vanb89}) and discussed more recently also by Mattila \& Meikle
(\cite{mattila}).
The first K-band (2.2$\mu$m) monitoring campaign was
attempted by Van Buren et al. (\cite{vanb94}).
They discovered SN1992bu by comparing
four near-IR images of NGC3690 (this SN was not confirmed
spectroscopically). More recently, a search in the K$'$-band was attempted by
Grossan et al. (\cite{grossan});
they failed in detecting any new SN and they
ascribe their negative result to the poor resolution of their camera.

We started a K$'$-band ($\rm 2.1\mu$m) monitoring campaign
of a sample of $\sim 35$ Luminous Infrared
Galaxies (LIRGs) aimed at detecting obscured SNe elusive
in optical surveys.
LIRGs are galaxies characterized by far-IR (FIR) luminosities
in excess of $\rm 10^{11}L_{\odot}$.
According to optical and IR studies the main energy source of this
class of objects seems to be starburst activity, at least
up to luminosities of $\rm \sim 10^{12.5}L_{\odot}$ (e.g.
Sanders \& Mirabel \cite{sanders}, Genzel et al. \cite{genzel}).
The contribution to
the IR luminosity from hidden active galactic nuclei (AGNs) is matter of
debate. However,
if most of their luminosity is powered by star formation,
then the inferred star formation activity would imply an average
 rate of about one SN
per year per object, within our sample (Mattila \& Meikle \cite{mattila}).

The survey, started in late 1999, was carried on with the
NTT-ESO, TNG\footnote{The Italian Galileo National Telescope}
and Kuiper/Steward\footnote{University of Arizona} telescopes.
In this paper we report the discovery of two infrared SNe: SN2001db
(Maiolino et al. \cite{maiolino01})
and SN1999gw (Cresci et al. \cite{cresci}), and
provide a preliminary estimate of the SN production in Luminous
Infrared Galaxies.
A more detailed description of our survey and more accurate estimate
of SN rate will be given
in a forthcoming paper (Mannucci et al. in prep.).

\section{SN2001db}

\subsection{Observations}

\subsubsection{Imaging}

The infrared imaging observations were obtained with SOFI, the near-IR
camera (Lidman et al. \cite{lidman})
 at the NTT ESO telescope, in the Ks band (see Tab.1).
NGC3256 was observed
for the first time on January 9, 2001, and subsequently three more
times: February~8, March~18, and April~1.
Each observation consisted of a series of 15 images, 60 seconds each,
alternated with 15 images sampling the near sky for an optimal
removal of the background. The total on-source integration time
was therefore 15 minutes per epoch, obtained under sub-arcsec seeing
conditions.
The data reduction was carried out with a
``home'' package developed by our team (Hunt et al. \cite{hunt94}).
The images obtained during the monitoring in 2001 were compared with
an archival image obtained with SOFI on November 28, 1999.

In Fig.1 we show the archival SOFI image of NGC3256 (left) and the
first image obtained in 2001 (right) where SN2001db has been detected.
The magnitude of the SN at the discovery epoch was
Ks=$16.03\pm 0.20$ \footnote{The difference with respect to the magnitude given
in Maiolino et al. (\cite{maiolino01})
is due to a better subtraction of the background
light.}. The SN is located at R.A.(J2000)~=~10h27m50s.4,
Decl.(J2000)~=~--43 54$'$ 21$''$, i.e. offset 5$''$.7 to the
West and 5$''$.7 to the South of the Ks nucleus of the
galaxy (uncertainty on the relative position about 0$''$.4).
In Tab.2 and in Fig.2 (upper panel) we give the Ks-band light
curve, both in terms of relative and absolute magnitude (assuming
a distance modulus of 33.06).

Photometric measurements performed on ESO archival images (Feb 1993)
and FORS images (Apr 2001, May 2001)
imply m$_V >$22 for the SN (Fig.2 lower panel).

\begin{figure*}[!]
\centering
\includegraphics[width=12truecm,angle=0]{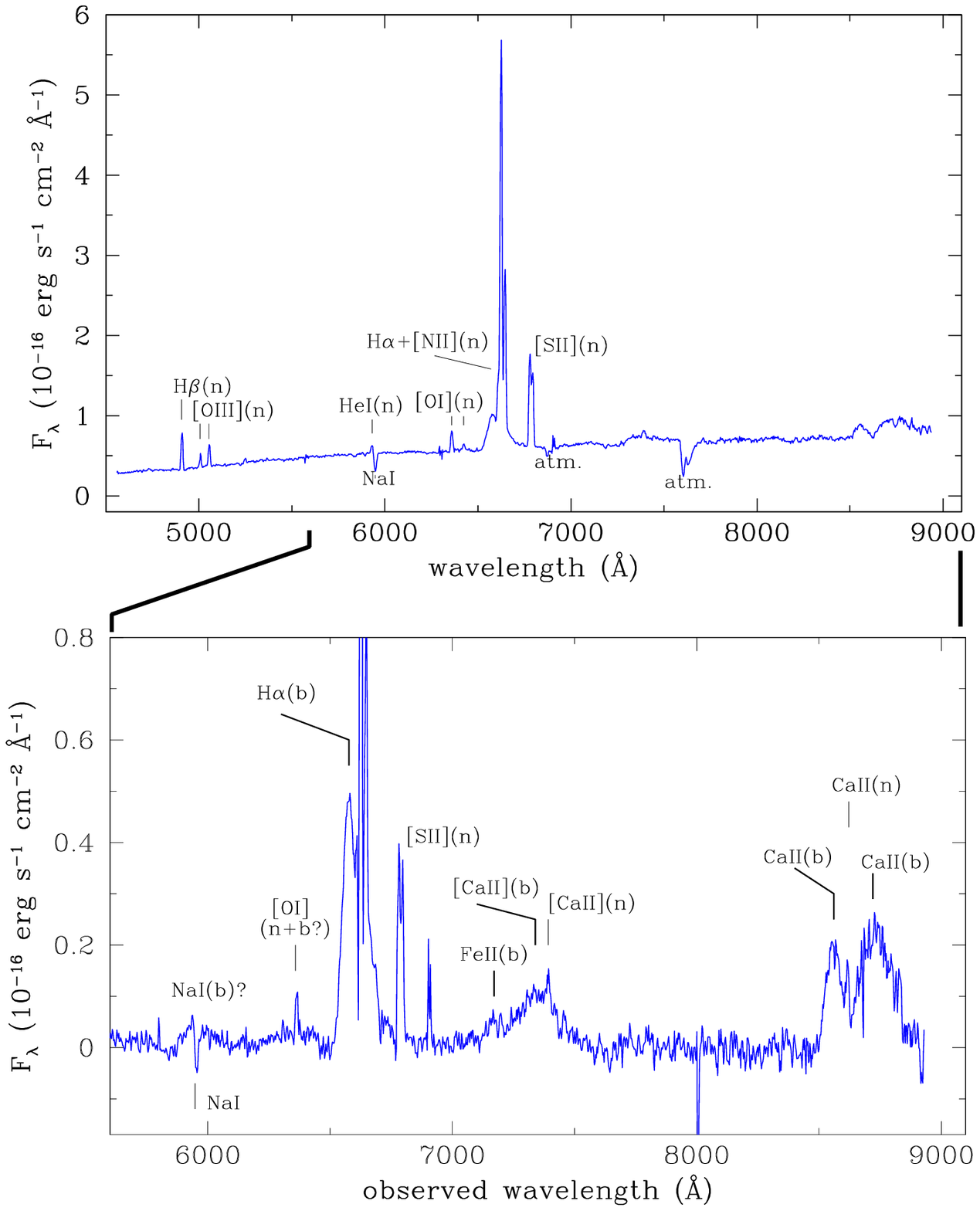}
\caption{{\sl Upper panel:} Optical spectrum of the SN obtained on May
16$^{th}$, 2001. Only the narrow emission lines and absorption lines
are marked.  {\sl Lower panel:} Optical spectrum after removing the
background galaxy light.  {\it n} and {\it b} refer to narrow and broad
components, respectively; the former most likely associated to the
background HII regions while the latter are associated to the SN.}
\end{figure*}

\begin{figure*}[htb!]
\centering
\includegraphics[width=12truecm,angle=0]{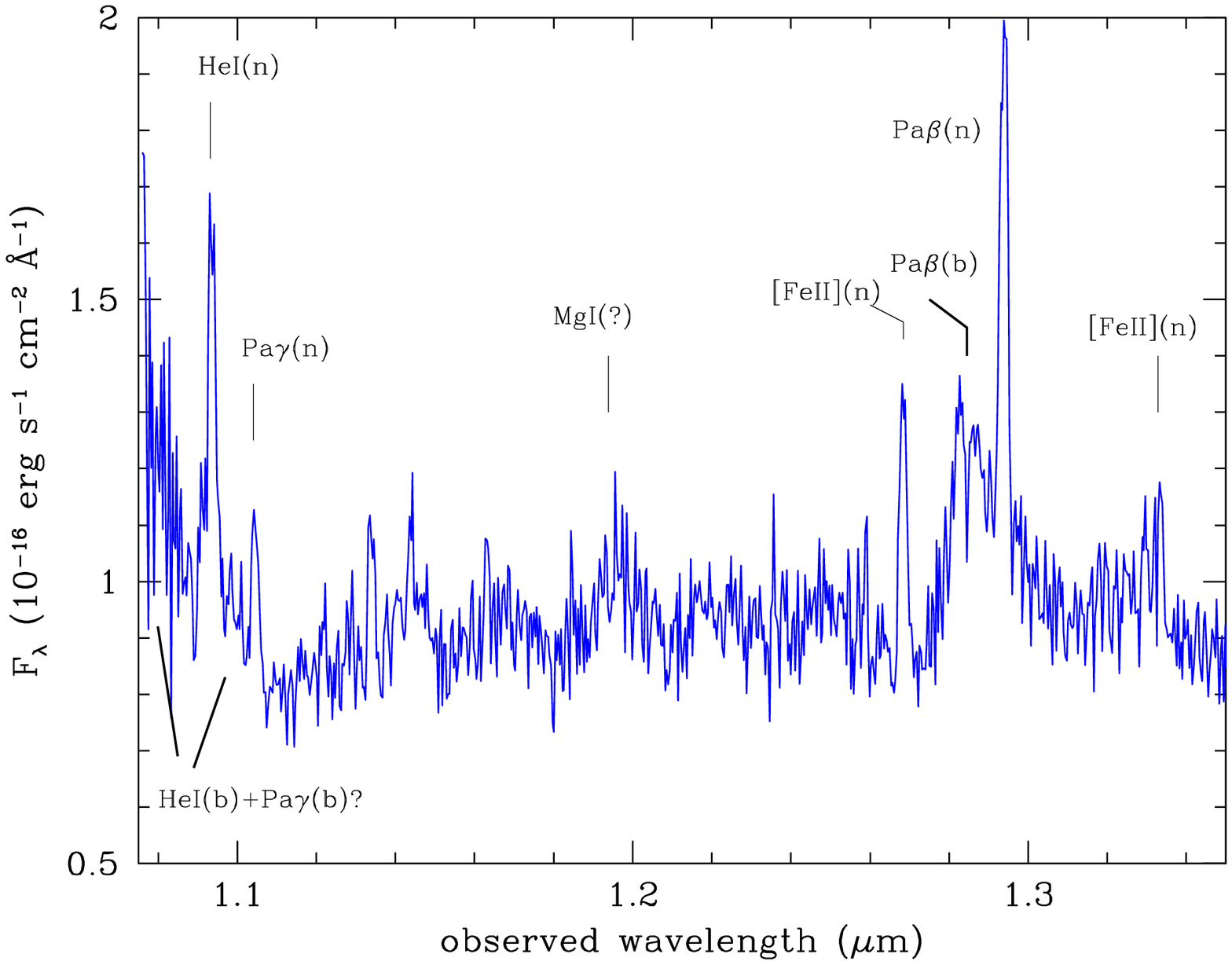}
\caption{Infrared spectrum obtained on April 21$^{st}$, 2001.
{\it n} and {\it b} refer to narrow and broad
components, respectively; the former most likely associated to the
background HII regions and SN remnants,
while the latter are associated to the SN2001db.}
\end{figure*}

\subsection{Optical spectroscopy}

The spectroscopic follow-up  was performed
both in the optical and in the infrared.

The optical spectrum was obtained with FORS1 (Szeifert \cite{szeifert})
at the ESO VLT-UT1 on May
16, 2001 (i.e. when the SN was at least 4 months old) with the grism
GRIS\_300V and a spectral resolution of 500. The 15m exposure was
split in two different integrations, to allow an optimum subtraction
of cosmic rays.
Background was removed by interpolating the spectrum above and
below the galaxy (after standard flat-fielding and bias subtraction).

Fig.3 (upper panel) shows the optical spectrum of the SN2001db.  There
are several narrow emission lines most of which are probably to
ascribe to HII regions and SNRs in the environment of SN2001db.  As
mentioned above, the continuum is not due to the SN2001db but is
dominated by background emission from the host galaxy.  The spectrum
clearly shows a broad component of H$\alpha$ (FWHM $\sim$5000 km/s)
which is a clear signature of the SN. The
signatures of the SN are better observed if the contribution of the
the underlying background continuum
is removed (by using as a template the contiguous regions of the galaxy
intercepted by the slit and properly re-scaled).
The resulting spectrum is shown in Fig.3 (lower panel).
The H$\alpha$ has a strongly asymmetric
profile, whose peak is blueshifted by about 2000 km/s with respect
to the parent galaxy (see Schlegel \cite{schlegel}
for a discussion on the blueshift),
but it also has
a prominent red tail, a profile similar to that observed in
type IIL and IIP SNe (Filippenko \cite{filippenko}).
No other SN signatures are found in the blue part
of the spectrum.
A summary of the properties of
all the broad lines associated to the SN is given in Tab.~2.
We note that the profile of the
broad component of [Ca II]~7324\AA \ and CaII~8662\AA \
is not as asymmetric as the other broad lines.
At the same redshift of the peak of the broad H$\alpha$ we have also detected
the FeII 7155\AA \ line.

\subsection{Infrared spectroscopy}

The infrared spectrum was obtained in the J band with ISAAC
(Cuby et al. \cite{cuby}) at the
ESO VLT-UT1 on April 21, 2001.
Observations were performed with the $1''$ slit and with the
low resolution grating (set at the 4$^{th}$ order), giving
a spectral resolution of 500.
The observations were obtained with
a set of single integrations of 30 seconds each (10 times 3 sec), by
moving the object along the slit to obtain an optimal sky subtraction,
for a total of 45 minutes of integration. Finally, the spectra obtained
at different
positions along the slit,
were subtracted from each other to remove the background,
then flat fielded, aligned, co-added and calibrated.
The atmospheric trasmission features were corrected as described
in Maiolino et al. (\cite{maiolino96}).

Fig.4 shows the infrared ISAAC spectrum extracted from an aperture of 1$''$.
Here the broad component of Pa$\beta$ is more prominent, relative to the
narrow component, with respect to H$\alpha$. The profile of the broad
component of Pa$\beta$ is nearly identical to that of H$\alpha$.
There are also indications of a
broad component of Pa$\gamma$ and HeI 1.0083$\mu$m, but
assessing the reality of these features is difficult because
it is at the edge
of the spectrum. Pa$\beta$ might be characterized by an absorption
on the blue side (i.e. a P-Cygni profile) which might be filled by
the [FeII] line. However, the flux of the latter
line is a factor of $\sim$3 higher than the companion [FeII]1.3206$\mu$m, as
expected by atomic constants, thus it is unlikely that [FeII]1.2567$\mu$m
is affected by a significant
absorption feature beneath its profile. Most likely
Pa$\beta$ does not have a P-Cygni signature,
similarly to H$\alpha$.
It is worth noting that the strong [FeII] emission relative
to Pa$\beta$ (narrow) indicates that the underlying emission is not simply
due to HII regions, but must be contributed significantly also
by SN remnants.

\begin{figure}[h!]
\centering
\includegraphics[width=\linewidth,angle=0]{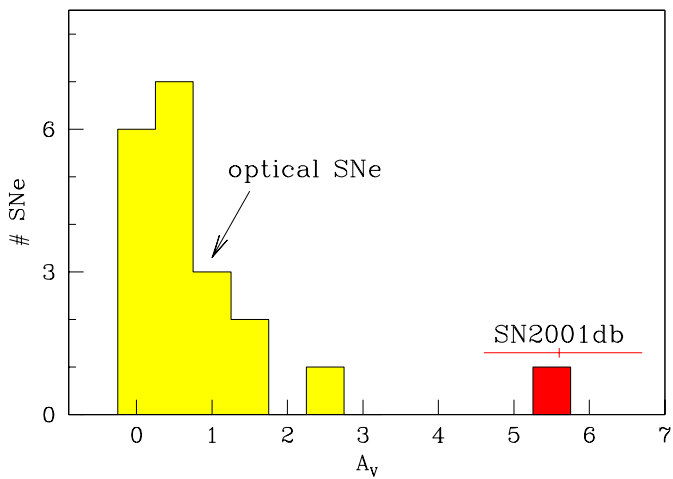}
\caption{
Distribution of extinction for the optically discovered SNe reported
in the literature (see text), compared with the extinction measured for
the infrared SN2001db.}
\end{figure}

\subsection{Extinction}

The Balmer decrement for the narrow components of H$\alpha$ and H$\beta$ is
$\sim 12$. This implies, for a Galactic extinction curve,
an equivalent {\it screen}
extinction $\rm A_V = 4.2$ mag. If dust is mixed with
 the emitting gas the
real optical depth is even
higher\footnote{Note that the ``pure'' {\it mixed}
case would give a maximum H$\alpha$/H$\beta$ ratio of $\sim 4.1$, implying that
at least part of the absorption is ascribed to a simple foreground screen
of dust.}.
The Na I interstellar absorption doublet at $\sim$5890\AA \ has an equivalent
width of 5.87 \AA .  However we note that due to the blending with the near
HeI $\lambda$5876\AA \ emission line, the equivalent width of Na D has been
underestimated. According to the relations of Barbon et al.
(\cite{barbon90}) and
Benetti (priv. comm.), and after
assuming a Galactic extinction curve, EW(Na D)$\geq$5.87\AA \ implies an
equivalent {\it screen} extinction $\rm A_V\geq3-4.9$ mag.

\begin{figure*}[htb!]
\centering
\includegraphics[width=12truecm,angle=0]{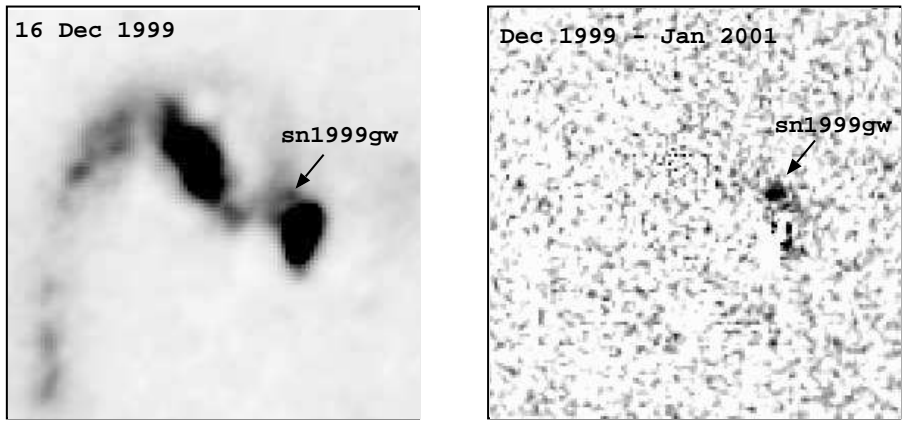}
\caption{Left: Ks-band image of UGC4881
obtained with ARNICA in December 1999.
Right: difference between the ARNICA image of December 1999 and the NICS
image of February 2001, after matching of the two PSF.
Both images are $35''\times 35''$ in size.
}
\end{figure*}

We can exploit the ratio of the broad component of Pa$\beta$ and
H$\alpha$ to constrain directly the reddening affecting the SN.  We
will assume that during 25 days elapsed between the infrared and
the optical spectrum the line flux has not changed significantly
(e.g. Danziger et al. \cite{danziger},
Xu et al. \cite{xu}, Benetti et al.
\cite{benetti});
afterwards we will discuss the case of rapid line variability.  The
observed ratio between the broad components of Pa$\beta$ and H$\alpha$
is 1.73.  The case B recombination gives an intrinsic ratio
Pa$\beta$/H$\alpha = 0.06$, which would imply an extinction of $\rm
A_V = 6.7$ mag. Yet, in type II SNe the intrinsic ratio
Pa$\beta$/H$\alpha$ appears to be lower than in the case B, although
only a few simultaneous observations of Pa$\beta$ and H$\alpha$ are
available for type II SNe.  The Pa$\beta$/H$\alpha$ ratio observed in
1987A was about 0.12 (Xu et al. \cite{xu}).  By interpolating the data
obtained by Fassia et al. (\cite{fassia00}) at different epochs
we obtain a similar ratio Pa$\beta$/H$\alpha
= 0.09$, once corrected for foreground reddening. If we conservatively
assume Pa$\beta$/H$\alpha =0.12$, the observed value in SN2001db gives
an extinction toward the SN of $\rm A_V = 5.3$ mag.

To discuss the quick variability case we assume that the hydrogen line flux
has decreased at the same rate of the continuum. In this case
the line flux cannot have changed by more than 40\%, which gives a lower
limit on the inferred optical extinction of $\rm A_V > 4.6$ mag or
$\rm A_V > 5.7$ mag for an intrinsic ratio of 0.12 and 0.06 (case B)
respectively.

Summarizing, we derive that the optical extinction
toward the SN is larger than $\rm A_V > 3$ (lower limit
inferred above from the Na D absorption and narrow Balmer decrement)
and most likely in the range 4.6 to 6.7 mag. In the following we assume
$\rm A_V \approx 5.6$ mag, with an uncertainty of about
1 mag.

Note that the upper limit on the V magnitude of the SN also provides
a constraint on the extinction when compared with the light curve
of SN templates (Fig.~2). With an extinction $\rm A_V < 2$ mag the
SN would have been detected in the optical image of April 17. This
lower limit on the extinction $\rm A_V > 2$ is fully consistent with
the value inferred above.

In this section we have adopted a Galactic
``standard'' extinction
curve, but there are indications that starbust galaxies might
have different extinction curves.
Calzetti et al. (\cite{calzetti})
estimated the extinction curves for a sample
of starburst galaxies. These extinction curves apply mostly
for the "mixed" case and, therefore, are probably inappropriate
for our SN. However, aware of this caveat,
we have also tried to estimate the extinction by using the
extinction curves given in Calzetti et al. (\cite{calzetti}) and
obtained (accounting also the uncertainties in the reddening curves)
a visual extinction ranging from 4.1 to 8.4 mag.

In the compilation given in Mattila \& Meikle (\cite{mattila})
and in Schmidt et al. (\cite{schmidt})
the extinction A$_V$ of core collapsed SNe is
generally lower than $\sim 1.5$ mag, with the exception of SN 1973R
for which Schmidt et al. estimate $\rm A_V = 2.7$ mag. The distribution
of extinctions in these two compilations is shown in Fig.~5.
If these compilations are representative, then
the extinction inferred for the IR SN 2001db
is probably the highest among the SNe so far discovered (Fig.~5).

\subsection{Supernova type and age}

The lack of any absorption blueward of the line peak (i.e. the typical
P-Cygni profile) may dis-favor a type IIP. However, since the
continuum is dominated by the background emission of the parent
galaxy, the lack of P-Cygni absorption is more difficult to
assess. Based on Fig.3 (lower panel), we cannot exclude the presence
of an absorption feature blueward of the H$\alpha$ with an EW less
than $\sim$ 2\AA .
 The shape of the broad hydrogen lines
(strongly asymmetric and with blueshifted peak) does not favor a type
IIn. The available photometric points also do not
discriminate between type IIL and IIP.  In Fig.2 we report the light
curves of SN1980k and SN1999em taken as representative of type IIL and
type IIP, respectively (data from Dwek et al. \cite{dwek},
Barbon et al. \cite{barbon82},
Hamuy et al. \cite{hamuy}). Both the optical and infrared light
curves have been extinguished by an A$_V =5.6$ (i.e. $\rm A_K=0.56$),
as obtained in the former section.  The infrared observations can be
roughly fit both with the light curve of SN1980k offset by 65 days,
and with the light curve of the SN1999em offset by 90 days and 1.0
mag. The average K-band light curve for type II SNe obtained by
Mattila \& Meikle (\cite{mattila}) is nearly identical to the SN1980k light
curve in Fig.2, but offset by 55 days.

It is interesting to note that, in any case, the
V-band magnitude of the SN was most likely fainter than 20 at
its maximum (see Fig.2)
and, therefore, it would have been missed by most of the optical SN search
programs (e.g. Richmond et al. \cite{richmond}).

\section{SN1999gw}

The infrared imaging observations were obtained with the
ARNICA near-IR camera (Hunt et al. \cite{hunt96})
at TNG telescope in the Ks band (see Tab.1). UGC4881 was observed for the first
time on December 16, 1999 and subsequently three more times: January 12 and
February 12, 2000 and on February 8, 2001 with the NICS camera (Baffa et al.
\cite{baffa}).

The observing strategy and data reduction were analogous to SN2001db.
The total on--source integration time for each epoch was 15 min.
In Fig.6 (left) we show the ARNICA image of UGC4881, where the SN was detected.
The right panel of Fig.6 shows the difference between the image of December
1999 and the image of February 2001, after properly matching the PSF of the
two images by means of the ISIS software (Alard \cite{alard}).
SN1999gw is clearly detected in emission.
Some residuals are observed on the nucleus south of the SN (i.e. UGC4881A),
where the steeply rising surface brightness prevents an accurate
subtraction of the PSF. The magnitude of the SN on December 1999
was Ks=$17.45\pm 0.20$. The SN was located at R.A.(J2000)~=~09h15m54.7s,
Decl(J2000)~=~+44~19$'$~55$''$,
being offset 3.5$''$ to the North of the Ks nucleus of UGC4881A.
In Tab.1 and in Fig.~7 we give
the Ks-band light curve (the absolute magnitude is derived by assuming a
distance modulus of 36.22).

In Fig.~7 we overplot the average light curve derived by Mattila \&
Meikle (\cite{mattila}), which fits fairly well
the photometric points of SN1999gw if the date of discovery is close to
the maximum.

\begin{figure}[h!]
\centering
\includegraphics[width=8truecm,angle=0]{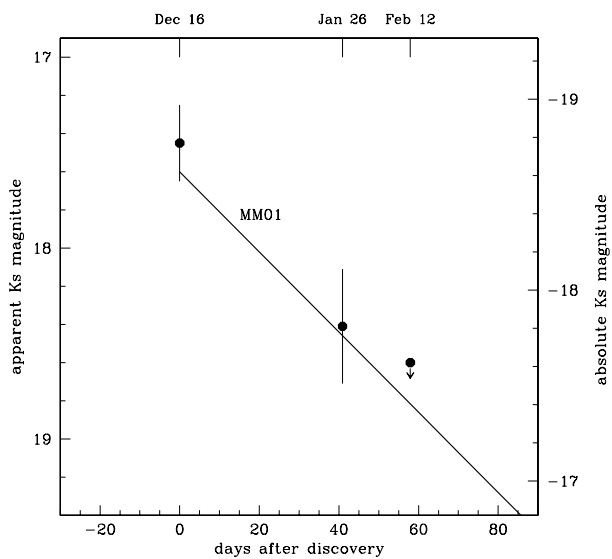}
\caption{Ks-band light curve of the SN1999gw.
The solid line is the average light curve obtained by Mattila and Meikle
(\cite{mattila}, MM01).}
\end{figure}

\section{The infrared SN rate}

Our monitoring program is still in progress. A final, detailed summary
of our survey and its implications on the SN rate
will be provided in a forthcoming paper (Mannucci et al. in
prep.). Here we compare
the number of SNe detected during our survey with the number of SNe
expected by applying the current estimates of SN rates, expressed
in SNu
(Cappellaro et al. \cite{cappellaro}), to our sample of galaxies.
So far, our program has detected 4 SNe. One of these is a type Ia
(SN1999gd, Li \cite{li}), which
was also detected in the optical. Another SN was a type II (SN2000bg,
Sato \& Li \cite{sato}),
also detected
in the optical. The other two SNe, SN1999gw and SN2001db, were
discussed in this paper.

By assuming the conversion from blue luminosity to SN rate (all types)
given in Cappellaro et al. (\cite{cappellaro}),
i.e. $\rm SNr \simeq 10^{-12}
(L_B/L_{\odot})~yr^{-1}$, we would have expected to detect $\sim 0.5$
SNe.  Since we have detected 4 SNe, we roughly infer a SN
production which is about an order of magnitude higher than estimated
by conventional optical surveys.

This high SN rate reflects both the higher extinction affecting
the B light of LIRGs (which is the normalizing factor of SNu)
and the enhanced star formation in
the galaxies of our sample. The enhanced SN rate obtained by our IR
survey partly reconciles the contractictory
results on starbursts obtained by optical surveys, which were discussed
in the Introduction.

Note that a SN rate higher than expected by the blue luminosity
is also obtained by the two SNe detected in the optical,
implying that the high SN rate inferred by our study is not only
due to obscuration. An important difference of our study,
with respect to previous surveys, is that our sample is characterized
by a much stronger starbursting activity; more
specifically, if the star formation is estimated through the far-IR
luminosity, the galaxies in our sample are characterized by a star
formation rate which is about one order of magnitude higher
than in the sample of Richmond et al. (\cite{richmond}).
Probably, the failure of previous studies in detecting an
enhanced SN rate in starburst galaxies is partly to ascribe also to
the selected sample, which consisted mostly of ``mild'' starbursts.

Altough higher than found in optical surveys, the SN rate measured
by us is still not as high as
expected by the large staformation rate inferred from the far-IR
luminosity of the galaxies in our sample.
Indeed, if most of the far-IR
luminosity is due to star formation, and we adopt the conversion
from L$_{FIR}$ to SN rate given in Mattila \& Meikle (\cite{mattila}),
i.e. $\rm SNr \simeq 2.7 \times 10^{-12} (L_{FIR}/L_{\odot})~yr^{-1}$,
we find that our survey has detected only about 20\% of the
expected SNe.

There are various possible scenarios which could explain the shortage
of SN detections. One possibility is that most SNe are so
embedded into the dust to be significantly obscured even in the near-IR.
More specifically, in order to reduce the expected number of SNe by
a factor of $\sim$5 (i.e. the fraction of missing SNe) they should
be, on average, fainter by 3.0 magnitudes in K, implying
that they should be absorbed by $\rm A_K > 3.0$, or $\rm A_V > 30$.
Another possibility is that
most SNe occur in the nucleus (i.e. within the central 2$''$); in this
case
our limited angular resolution would have prevented disentangleing
them from the peaked nuclear surface brightness of these galaxies, even
when the PSF between multi-epoch images are optimally matched.
In support to this scenario recent mid-IR studies have found growing
evidence that a large fraction of the
starburst activity occurs in the nuclear region (Soifer et al.
\cite{soifer}).
The possibility that most SNe occur in the nuclear region
can be tested through near-IR monitoring
from space (e.g. NICMOS on HST), which would provide higher angular resolution
and more stable PSF. Alternatively, nuclear SNe could be disentangled
spectroscopically through the detection of the broad hydrogen lines
characteristics of type II SNe, by periodically monitoring the (near-IR)
spectra of the nuclei of the galaxies in our sample. The discovery of
a nuclear SN by Aretxaga et al. (\cite{aretxaga})
through the detection of broad
hydrogen lines has shown the feasibility of the latter approach in
the optical.

Finally, the assumption that the high far-IR luminosity is
tracing an enhanced star formation rate might be incorrect.
In particular, obscured
AGNs may contribute substantially to the IR luminosity
of these galaxies (Sanders \& Mirabel \cite{sanders}). Note that heavily
obscured AGNs may be present even if not identified
spectroscopically and may be totally obscured even
in the X-rays (Marconi et al. \cite{marconi},
Maiolino et al. \cite{maiolino98}).

\section{Conclusions}

We have presented the discovery of SN~1999gw and SN~2001db obtained
within the framework of a near-IR ($\lambda \sim 2.1\mu$m) search for
obscured SNe in Luminous Infrared Galaxies.  The spectroscopic
follow-up of SN2001db indicates that this was a type II SN, extinguished
by A$_V \approx 4.6-6.7$.  Interesting enough
this SN, the first one to be discovered in IR and spectroscopically confirmed,
 would have been missed by ordinary optical surveys.

Preliminary estimates of the SN rate indicate that the SN production
in Luminous Infrared Galaxies is
$\rm \sim 10^{-11} (L_B/L_{\odot})~yr^{-1}$,
about an order of magnitude higher than
expected by the relation between L$_B$ and SN rate found in optical
studies.  This result confirms that optical surveys miss a significant
fraction of the SNe, especially in obscured systems such as starburst
galaxies. However, simple first order estimates, based on the far-IR
luminosity, indicate that our near-IR survey is still largely
incomplete, and that $\sim$ 80\% of the SNe are missed.  Some possible
explanations for this issue are: 1) most SN are more obscured than
$\rm A_V>30$ mag; 2) the number of expected SNe as inferred from the
L$_{FIR}$ is overestimated because hidden AGNs dominated the
luminosity; 3) most of the SNe occur in the nuclear region and the
limited angular resolution prevents their discovery.

\begin{acknowledgements}
We are grateful to S. Mattila, P. Meikle and N. Panagia for useful comments.
The near-IR survey partly presented in this paper is the result of the
efforts of a larger collaboration which also includes
V. Ivanov, A. Alonso-Herrero and N. Nagar.
We thank the ESO staff, on La Silla and on Paranal, as well as the TNG staff
 for their help
during the observations. We thank the ESO General Director for
allocating Director Discretionary Time to this project.
 M. Della Valle and R. Maiolino
are grateful to the Paranal and La Silla
ESO observatories and to the ESO Santiago Science Office
for their kind hospitality during this research program, within the
framework of the Visiting Scientist Program.
This work was partially supported by the
the Italian Space Agency (ASI).
\end{acknowledgements}

\begin{table*}
\begin{center}
\begin{tabular}{lcccc}
\hline
\multicolumn{4}{c}{SN2001db}\\
Date & Tel./Instr. & type & band/$\Delta \lambda$ & mag \\
09 Jan 2001 & NTT/SOFI & imag. & Ks & 16.03$\pm$0.20\\
08 Feb 2001 & NTT/SOFI & imag. & Ks & 16.93$\pm$0.20\\
18 Mar 2001 & NTT/SOFI & imag. & Ks & 17.33$\pm$0.25\\
01 Apr 2001 & NTT/SOFI & imag. & Ks & 17.74$\pm$0.30\\
17 Apr 2001 & VLT/FORS1 & imag. & open & V$>$22\\
21 Apr 2001 & VLT/ISAAC & spec. & 1.08--1.35$\mu$m & \\
16 May 2001 & VLT/FORS1 & spec. & 4500--9100\AA  & \\
16 May 2001 & VLT/FORS1 & imag. & open & V$>$22\\
\hline
\multicolumn{4}{c}{SN1999gw}\\
Date & Tel./Instr. & type & band/$\Delta \lambda$ \\
16 Dec 1999 & TNG/ARNICA & imag. & Ks & 17.45$\pm$0.20\\
26 Jan 2000 & TNG/ARNICA & imag. & Ks & 18.41$\pm$0.30\\
12 Feb 2000 & TNG/ARNICA & imag. & Ks & $>$18.6 \\
08 Feb 2001 & TNG/NICS   & imag. & Ks & $>$18.6 \\
\hline
\end{tabular}
\caption{Log of observations and photometry of the two SNe discussed
in this paper.}
\end{center}
\end{table*}

\begin{table*}
\begin{center}
\begin{tabular}{llccc}
Line      & $\lambda _{rest}$&Flux       & FWHM$^a$      & V$_{peak}^{b}$ \\
         &     &  $\rm (10^{-15}erg~s^{-1}cm^{-2})$& $\rm (km~s^{-1})$
                                                   & $\rm (km~s^{-1})$ \\
\hline
NaI$^c$    & 5893\AA          &0.34$\pm$0.07 & 6540$\pm$1000 &    340$\pm$800 \\
$[$OI$]^c$  & 6300\AA          &0.10$\pm$0.03  & 9500$\pm$2000 &    320$\pm$900 \\
H$\alpha$ & 6562\AA          &4.40$\pm$0.40  & 5200$\pm$600  & --2140$\pm$250 \\
FeII      & 7155\AA          &0.17$\pm$0.05  & --$^d$        & --2290$\pm$500 \\
$[$CaII$]$    & 7324\AA          &1.05$\pm$0.1   & 8416$\pm$500  & --1820$\pm$500 \\
CaII      & 8542\AA          &0.94$\pm$0.1   & 3000$\pm$800  & --2310$\pm$300 \\
CaII      & 8662\AA          &1.50$\pm$0.15  & 6500$\pm$700  & --370$\pm$300 \\
Pa$\beta$ &12818\AA          &7.61$\pm$0.7   & 4900$\pm$700  & --2570$\pm$300 \\
\hline
\end{tabular}
\caption{Broad lines detected and ascribed to the SN. 
Pa$\beta$ was observed on April 21, 2001, while all the optical lines
were observed on May 16, 2001. Quoted errors are at 1$\sigma$.
Notes:
$^a$ Full width half maximum in km s$^{-1}$.
$^b$ Velocity shift of the line peak with respect to
the rest frame velocity of the galaxy, in km s$^{-1}$.
$^c$ Marginal detection.
$^d$ Strongly blend with CaII.}
\end{center}
\end{table*}

\end{document}